\documentclass[aps,prd,twocolumn,groupedaddress]{revtex4-1}
\usepackage{graphicx}
\usepackage{amsmath}
\def\drv{ {\rm d}\! }

\def\dd{ {\rm \delta}\! }

\def\sim{ \simeq }

\def\H{{\mathcal H}}

\newcommand{\be}{\begin{equation}}
\newcommand{\ee}{\end{equation}}
\newcommand{\bea}{\begin{eqnarray}}
\newcommand{\eea}{\end{eqnarray}}

\begin{document}
\title{The thermodynamic evolution of the cosmological event horizon}
\author{Scott Funkhouser}
\email[]{scott.funkhouser@gmail.com}
\noaffiliation

\date{\today}

\begin{abstract}
By manipulating the integral expression for the proper radius $R_e$ of the cosmological event horizon (CEH) in a Friedmann-Robertson-Walker (FRW) universe, we obtain an analytical expression for the change $\dd R_e$ in response to a uniform fluctuation $\dd\rho$ in the average cosmic background density $\rho$.  We stipulate that the fluctuation arises within a vanishing interval of proper time, during which the CEH is approximately stationary, and evolves subsequently such that $\dd\rho/\rho$ is constant.  The respective variations $2\pi R_e \dd R_e$ and $\dd E_e$ in the horizon entropy $S_e$ and enclosed energy $E_e$ should be therefore related through the cosmological Clausius relation.  
In that manner we find that the temperature $T_e$ of the CEH at an arbitrary time in a flat FRW universe is $E_e/S_e$, which recovers asymptotically the usual static de Sitter temperature.
Furthermore, it is proven that during radiation-dominance and in late times the CEH conforms to the fully dynamical First Law $T_e \drv S_e = P\drv V_e - \drv E_e$, where $V_e$ is the enclosed volume and $P$ is the average cosmic pressure.
\end{abstract}
\maketitle

\section{Introduction \label{Intr}}
Following from the seminal work of Gibbons and Hawking \cite{GH1977}, a generalized prescription has been developed to describe the thermodynamic properties of both cosmological event horizons and the event horizons of black holes, for a broad class of static metrics of the form 
\be
\drv s ^2 = f \drv t^2 - f^{-1}\drv r^2  - r^2 \drv \Omega ^2 ,
\label{EqM1}
\ee
given in the usual notation and in Planck units, which are used throughout this work \cite{Pad2002}, \cite{Pad2005}.  
The radial function $f=f(r)$ determines the characteristics of the corresponding spacetime.  
Each finite, positive radial coordinate $r=R_j$ that is a simple solution of $f(r)=0$ coincides with a spherical event horizon.  The surface gravity $\kappa _j$ on the $j$-th horizon is given by
\be
\kappa _j = \frac{1}{2} f' (r) \bigg| _{r=R_j}
\label{EqKj}
\ee
and the temperature $T_j$ and entropy $S_j$ of the horizon are, respectively, \cite{Pad2002}, \cite{Pad2005}
\be
T_j = \frac{| \kappa _j|}{2 \pi}
\label{EqTj}
\ee
and
\be
S_j = \pi R _j ^2.
\label{EqSj}
\ee

Event horizons in static spacetimes are generally subject to a thermodynamic rule that is analogous to the conventional Clausius relation.  
Consider a large conventional system, with temperature $T$, entropy $S$ and pressure $p$, occupying a spatial volume $V$ that contains a total energy $E$.  Suppose that the system is subject to an infinitesimal thermodynamic transition during which $S$, $E$ and $V$ are allowed to evolve freely.  From the First Law of Thermodynamics (FLT), we have  
\be
T \drv S = \drv E + p \drv V .
\label{EqFLT}
\ee  
Suppose, alternatively, that an infinitesimal, reversible transfer of heat between the system and the adjacent environment occurs, thereby changing $E$ by some quantity $\dd Q$.
It follows from Eq.(\ref{EqFLT}), with $\drv V=0$, that the corresponding infinitesimal change $\dd S$ in the entropy is given by 
\be
\dd S \simeq \frac{\dd Q}{T} ,
\label{EqCR}
\ee
which is the relationship by which Clausius defined entropy.

Now let us consider a system consisting of a spherical causal horizon with proper radius $R$, entropy $S$ and temperature $T$.  Suppose that, as the consequence of a natural thermodynamic fluctuation, an infinitesimal quantity of energy crosses the horizon, changing the total enclosed energy $E$ by $\dd E$.  Let the fluctuation transpire rapidly enough to ensure that $R$, $E$ and $S$ remain approximately constant but for the effects of the fluctuation.
As such, the horizon is quasi-stationary with respect to the fluctuation and is therefore subject to the cosmological analogue of the conventional Clausius relation.  Specifically, the FLT for stationary event horizons specifies that \cite{Cai2002a}, \cite{Cai2002b}
\be
T \dd S \simeq -\dd E .
\label{EqFLTsch}
\ee
In the case of a stationary black hole horizon, the appropriate form of the FLT would be $T\dd S \simeq \dd E$.  
The reversal of the sign of $\dd E$ in Eq.(\ref{EqFLTsch}) with respect to the conventional Clausius relation represents an important physical distinction between cosmological horizons and most other thermodynamic systems.
Whereas the entropy of a stationary black hole increases as the enclosed energy increases, the entropy of a cosmological event horizon in a static spacetime increases as the enclosed energy {\it decreases} and the energy of the {\it exterior region} increases \cite{GH1977}.

An instructive example is the Schwarzschild-de Sitter (SdS) metric, which is of the form Eq.(\ref{EqM1}) with $f$ equal to  
\be
f_{dS} \equiv 1 - \frac{2m}{r} - \frac{\Lambda r^2}{3},
\ee
where $m$ is the mass of the centrally located spherical body, nominally representing a black hole, and $\Lambda$ is the cosmological constant.
The two non-negative solutions of $f_{dS}(r)=0$, defined here as $R_1$ and $R_2$, correspond respectively to the black hole event horizon and the cosmological event horizon (CEH).  Note that $R_1$ is never larger than $R_2$.
As $\Lambda$ vanishes, $R_1$ approaches the usual Schwarzschild radius $2m$.
Conversely, as $m$ vanishes $R_2$ approaches the de Sitter radius
\be
R_{\Lambda} = \left( \frac{3}{\Lambda} \right)^{1/2} .
\ee

If $m\ll R_{\Lambda}$ then
\be
R_2 \simeq R_{\Lambda} - m
\label{EqR2}
\ee
and the corresponding entropy and temperature are, respectively,
\be
S_2 \simeq \pi R_{\Lambda} ^2 - 2 \pi R_{\Lambda}  m
\label{EqS2}
\ee
and
\be
T_2 \simeq \frac{1}{2\pi R_{\Lambda} } -  \frac{m }{\pi R_{\Lambda}^2 } .
\label{EqT2}
\ee
The dS entropy and temperature are recovered from Eqs.(\ref{EqS2}) and (\ref{EqT2}) by putting $m=0$.  Suppose that some quantity $\dd m$ of energy were to escape from the central mass and migrate across the CEH.  According to Eq.(\ref{EqFLTsch}), with $T_2$ given by Eq.(\ref{EqT2}), the corresponding variation $\dd S_2$ must be 
\be
\dd S_2 \simeq -2\pi R_2 \dd m,
\ee
which is consistent with the result obtained directly by varying $m$ in Eq.(\ref{EqS2}) \cite{GH1977}. 

The thermodynamic properties of stationary horizons originate ultimately in the field equations of general relativity \cite{Jac2003}.
Insofar as the field equations are inherently thermodynamic, it is reasonable to expect that the basic thermodynamic behaviors attributed to stationary horizons should manifest generally, though perhaps in varying forms, in all relativistic causal horizons, including cosmological horizons associated with dynamical spacetimes.  
In principle we may probe the thermodynamic behaviors of a given dynamical horizon by analyzing the response of the horizon to an infinitesimal fluctuation in some system parameter.
Moreover, it is generally possible to consider a fluctuation that occurs rapidly, such that the horizon is quasi-stationary with respect to the fluctuation.  
In that manner we may determine the temperature of a dynamical horizon.
Specifically, if the horizon entropy varies by some quantity $\dd S$ in response to an infinitesimal change $\dd E$ in the enclosed energy, which transpires rapidly while such that the horizon is quasi-stationary, then the temperature of the horizon should be given by $\dd E/\dd S$ in accordance with Eq.(\ref{EqFLTsch}).
 
The cosmological apparent horizon (CAH) in a FRW universe is a centrally important example of a dynamical horizon whose temperature may be determined according to the prescription described in the previous paragraph \cite{AC2007}, \cite{CK2005}.
Consider a spacetime characterized by the FRW metric, which may be expressed as
\be
\drv s ^2 = \drv t^2 - a^2 \left( \frac{\drv r^2}{1-kr^2}  + r^2 \drv \Omega ^2 \right),
\label{EqFRW}
\ee
where $t$ is the proper time, $a$ is the cosmic scale factor and $k$ is the curvature parameter. 
At any given moment there exists a cosmological apparent horizon (CAH) whose proper radius $R_a$ is given by
\be
R_a = \left( H^2 + \frac{k}{a^2}   \right)^{-1/2} ,
\label{EqRa}
\ee
where
\be
H \equiv \frac{ \dot{a} }{a} = \left( \frac{8\pi}{3} \rho - \frac{k}{a^2}\right)^{1/2}
\label{EqH}
\ee
is the Hubble parameter, with the {\it overdot} signifying $\drv /\drv t$. 
The total energy contained in the volume $V_a$ enclosed by the CAH is $E_a = \rho V_a $, and its full differential is therefore $\drv E_a = \drv V_a \rho + V_a \drv \rho$. 

In order to determine the temperature of the CAH we may introduce a uniform fluctuation $\dd \rho$ such that the usual average cosmic background density at some arbitrary proper time $t_v$ becomes $\rho(t_v) + \dd \rho$.
Let the fluctuation arise within some vanishing interval $[t_v\!-\!\dd t, t_v]$ during which the CAH is quasi-stationary and $\rho$ is otherwise constant. 
The fluctuation therefore effects a nearly instantaneous change in energy $\dd E_a \simeq V_a \dd \rho$, which represents a transfer of energy while the CAH is quasi-stationary.
Following the standard formulation, let the entropy of the CAH be given by
\be
S_a = \frac{A_a}{4} ,
\label{EqSA}
\ee
where $A_a$ is the surface area of the CAH.  
With $\dd S_a \simeq 2\pi R_a \dd R_a$, and the limit as $\dd \rho$ and $\dd R_a$ become differentials, Eq.(\ref{EqFLTsch}) implies that the temperature $T_a$ of the CAH is given by
\be
T_a = -\frac{V_a }{2\pi R_a } \frac{\drv \rho}{\drv R_a} .
\label{EqTa0} 
\ee
Finally, it follows from Eq.(\ref{EqRa}) that 
\be
\frac{\drv R_a}{\drv \rho} = -V_a ,
\label{EqdRadrho}
\ee
and we therefore have
\be
T_a =  \frac{1}{2\pi R_a } ,
\label{EqTa}
\ee
which is the expected result \cite{AC2007}, \cite{CK2005}.

Presumably there exists a more general thermodynamic rule, analogous to the FLT, which describes the behaviors of dynamical horizons and recovers the stationary FLT as a special case.
The Unified First Law of Thermodynamics (UFLT), which was derived in the context of dynamical black holes, represents one form of such a rule \cite{Hay1998}, \cite{HMA1999}.
Suppose that a dynamical boundary, with surface area $A$, executes an infinitesimal thermodynamic transition in which the interior volume $V$ and the enclosed energy $E$ are free to evolve mutually.  According to the UFLT we have
\be
\drv E =  A \Psi + W \drv V ,
\label{EqUFLT}
\ee
where $\Psi$ is the associated energy-supply vector and $W$ is the work density \cite{Hay1998}.  

In the case of a spherical boundary with proper radius $R$ in a FRW universe, $\Psi$ and $W$ are respectively given by
\be
\Psi = \frac{1}{2}(\rho + P)\left( \drv R - \frac{R \drv a}{a} -HR \drv t \right)
\label{EqPsi}
\ee
and
\be
W = \frac{1}{2}(\rho - P) ,
\label{EqW}
\ee
where $P$ is the average cosmological pressure \cite{CK2005}.  It follows directly from Eqs.(\ref{EqUFLT}) through (\ref{EqW}) that
\be
\dot{E} = 4\pi R^2  \rho \dot{R} - 4\pi HR^3 (\rho + P).
\label{EqEdotR}
\ee
With a substitution from 
\be
\dot{\rho} = - 3H(\rho + P),
\label{Eqdrhodt}
\ee
which follows from the Friedmann equations, the right side of Eq.(\ref{EqEdotR}) immediately produces  $\dot{V} \rho  + V\dot{\rho}$, which is precisely the full derivative of $E$ with respect to $t$.  The field equations guarantee, therefore, that all co-moving spherical boundaries in a FRW universe conform to the UFLT. 

Notwithstanding its significance, the UFLT is not explicitly a function of temperature or entropy, and additional information is generally necessary in order to obtain a true analogue of the FLT from the UFLT.
Moreover, the UFLT never amounts to more than an explicit statement of $\dot{E}$ in the context of a FRW universe, and any relationship it produces must be already implied by the Friedmann equations, independently of the UFLT.

For instance, in the case of an effectively instantaneous fluctuation of energy across the CAH in a FRW universe, the UFLT implies that 
\be
\drv E _a = \frac{\tilde{\kappa}_a}{2\pi} \frac{\drv A_a}{4},
\label{EqUfc}
\ee
where $\tilde{\kappa} _a =  -1/R_a$ represents the surface gravity $\kappa _a$ evaluated with $\dot{R}_a =0$ \cite{CK2005}.  
Eq.(\ref{EqUfc}) may be obtained from the UFLT by expressing $A \Psi$ in the alternate form 
\be
A \Psi = \frac{\kappa }{ 2\pi} \frac{\drv A}{4} + R \drv \left(\frac{E}{R} \right).
\label{EqUfc2}
\ee
In the case of the CAH the term $R \drv (E/R)$ vanishes, and, with $\drv V=0$ and $\dot{R}_a = 0$, Eq.(\ref{EqUFLT}) therefore leads directly to Eq.(\ref{EqUfc}) \cite{CK2005}.
It follows from the FLT for stationary cosmological horizons that the right side of Eq.(\ref{EqUfc}) is equivalent to $-T_a \drv S_a$.  In accordance with Eq.(\ref{EqTj}), the term $-\tilde{\kappa}_a$ is identified as $2\pi T_a$, which implies that $S_a = A_a/4$ \cite{CK2005}.

Although the determination of $S_a$ from the UFLT is noteworthy, the same result may be obtained directly from Eq.(\ref{EqdRadrho}), which implies 
\be
\drv R_a = -V_a \drv \rho .
\label{EqUfc3}
\ee
The right side of Eq.(\ref{EqUfc3}) is equivalent to $-\drv E$ in the case where $\drv V_a =0$, which implies that the left side is equivalent to $T_a \drv S_a$.  With $T_a$ identified {\it a priori} from $\tilde{\kappa} _a$ as before, we readily infer from Eq.(\ref{EqUfc3}) that $\drv S_a = 2 \pi R_a \drv R_a$, and thus $S_a = A_a/4$.

Whereas the basic thermodynamic behaviors of the CAH are implicit to the Friedmann equations, the putative corresponding behaviors for the dynamical CEH are not so readily identified.  It is somewhat ironic that the field of cosmic holography emerged from the study of the stationary CEH, but the CAH has proven to be the most natural holographic horizon, at least in the current understanding\cite{Smoot2010}, \cite{GW2007}, \cite{BR2000}.
Although the relative prominence of the CAH is appropriate, the putative thermodynamic behaviors of other cosmological horizons may be important in forming a complete understanding of cosmic holography.  In particular, given that cosmological horizons in static spacetimes are subject to the FLT for stationary horizons, it is reasonable to expect that the CEH in a FRW universe is subject to an analogous rule.
After all, the cosmological horizon in a SdS or dS space is essentially the same entity as the dynamical CEH in a FRW universe, and the CEH and CAH converge in an asymptotically dS space.

The purpose of this present work is demonstrate that, with the properly defined temperature, the CEH does, in fact, conform to the FLT during the vacuum-dominated de Sitter era and also in a radiation-dominated FRW universe.
The temperature $T_e$ of the CEH is derived in the following Section by assessing the response to a variation in the enclosed energy that occurs rapidly, such that the horizon is quasi-stationary.  
The analysis in Section \ref{FLTE} demonstrates that, with the temperature obtained in Section \ref{Temp}, the thermodynamic evolution of the CEH is all but precisely described by a form of the FLT that is directly analogous to the conventional expression, but with the sign of the associated energy differential reversed in accordance with the FLT for stationary horizons.  
Section \ref{Disc} contains a discussion of the results.

\section{The dynamical temperature of the cosmological event horizon \label{Temp}}
Consider a spacetime characterized by a metric of the form Eq.(\ref{EqFRW}) which contains an ideal fluid consisting of relativistic, radiation-like particles and cold, dust-like matter.  Let the spacetime be also endowed with a positive cosmological constant.  The total density at any given time may be therefore itemized according to
\be
\rho = \epsilon + \mu + \nu ,
\label{Eqrho}
\ee
where $\epsilon$, $\mu$ and $\nu$ represent respectively the contributions from relativistic particles, cold dust and the cosmological vacuum.  The average cosmic pressure is accordingly
\be
P = \frac{\epsilon}{3} - \nu.
\ee
For convenience, let us define the equation-of-state parameter $w=w(t)$ such that 
\be
P = w \rho.
\label{EqEOS}
\ee
In addition to the vacuum-density
\be
\rho _{\Lambda} = \frac{\Lambda}{8 \pi}
\ee
associated with the cosmological constant, the vacuum-density $\rho _{\phi}$ associated with an inflationary scalar field $\phi$ may contribute to $\nu$.

Given that $\Lambda > 0$, there exists at all times a CEH with a finite proper radius 
\be
R_e = a \int _a ^{\infty} { \frac{\drv \alpha}{\alpha ^2 H (\alpha)} } ,
\label{EqRe}
\ee
where $H(a)=H$ represents the parameterization of $H$ in terms of the scale factor, or the corresponding variable of integration, $\alpha$.  Although the right side of Eq.(\ref{EqRe}) is parameterized in terms of $a$, it is convenient for the present purposes to regard $R_e$ as being implicitly a function of $t$.  It follows from Eq.(\ref{EqRe}) that
\be
\dot{R}_e = \frac{\drv R_e}{\drv a} \dot{a}= HR_e - 1.
\label{EqRedot}
\ee
The total energy enclosed within the horizon is $E_e = \rho V_e$, where $V_e =4\pi R_e ^3 /3$.  
We assume throughout the following that the entropy $S_e$ of the CEH is given by
\be
S_e = \pi R_e ^2 ,
\label{EqSE}
\ee
throughout all cosmological ages considered here.  The attribution of the entropy in Eq.(\ref{EqSE}) to a dynamical CEH is well motivated and produces physically consistent results \cite{Dav1987} \cite{Dav1988a}, \cite{Dav1988b}, \cite{PS1989}, \cite{Pav1990}, \cite{Bru2000}, \cite{IP2006}.

Analogously to the procedure used to identify the temperature of the CAH, let us determine the temperature $T_e$ of the CEH in a FRW universe by introducing an infinitesimal uniform perturbation $\dd \rho$ at some time $t_p$ such that the average cosmic background density becomes $\rho(t_p) + \dd \rho$.  Note that $\rho$ is the original background density in the absence of the perturbation.   
Let the magnitude of the perturbation increase rapidly from $0$ at time $t_v\!-\!\dd t$ to its characteristic magnitude $|\dd \rho|$ at time $t_v$, where $\dd t$ is small enough to ensure that the CEH is quasi-stationary and $\rho$ is otherwise nearly constant during the interval $[t_v\!-\!\dd t, t_v]$.   
The perturbation changes the enclosed energy at $t=t_v$ by $\dd E \simeq V_e \dd \rho$.
Let $\dd R_e$ represent the corresponding variation in the proper radius of the CEH in response to $\dd \rho$, measured with respect to $R_e(t_v)$.  
As $\dd t \dot{R}_e$ is negligible by design, $\dd R_e$ is approximately the total change in $R_e$ during the interval $[t_v\!-\!\dd t, t_v]$.
The perturbation thus changes the horizon entropy at time $t=t_v$ by $\dd S_e \simeq 2\pi R_e(t_v) \dd R_e$.
It follows from Eq.(\ref{EqFLTsch}) that $T_e$ is given by
\be
T_e \simeq -\frac{V_e }{2 \pi R_e} \frac{\dd \rho}{\dd R_e} .
\label{EqTe0}
\ee
Because $t_v$ is arbitrary it has not been shown explicitly in Eq.(\ref{EqTe0}), which is meant to be general.

In contrast to $R_a$, which is defined completely in terms of contemporaneous parameters, $R_e$ is determined by the manner in which the Hubble parameter evolves throughout all future times.  
In order to determine $\dd R_e$ it is therefore necessary to address the behavior of the perturbation throughout all future times.  
We may introduce time-dependence to the perturbation most conveniently by defining $f=f\!(t)$ such that the total background density at any time $t\ge t_v$ is given by $\rho + f \dd\rho$.  
We therefore have $f\!(t_v)=1$.  For all later times, $f$ is determined completely by a basic condition necessary for physical consistency.  
In order for the perturbation to be a physically meaningful probe it must be designed such that the universe evolves in the natural manner for all times after $t_v$.
We require that the additional contribution $f\dd\rho$ to the background density must evolve naturally, in the same manner as $\rho$, and hence
\be
\frac{f\dd \rho}{\rho} = \text{ {\rm const.} }
\label{Eqconst}
\ee

Let us now incorporate the time-dependent perturbation into Eq.(\ref{EqRe}), thereby obtaining an expression for $\dd R_e$.  
With $\rho$ replaced by $\rho+f\dd\rho$ the Hubble parameter for all $t\ge t_v$ becomes
\be
\left( H^2 + \frac{8\pi }{3} f\dd \rho \right)^{1/2} 
\simeq
H + \frac{4\pi f\dd \rho }{3H} .
\label{EqH2}
\ee
We accordingly define  
\be
\dd H \equiv \frac{4\pi f\dd \rho }{3H}  .
\label{EqdH}
\ee
At time $t_v$ the proper radius of the CEH may be expressed as
\be
R_e(t_v) + \dd R_e \simeq 
a \int _a ^{\infty} { \frac{\drv \alpha}{\alpha ^2  H \left(1 + \frac{\dd H}{H} \right)  } } .
\label{EqRedd}
\ee
A first-order expansion of the integrand in Eq.(\ref{EqRedd}) produces 
\be
\dd R_e \simeq - \frac{4\pi}{3} a \int _a ^{\infty} { \frac{f\dd \rho }{\alpha ^2  H^3 } \drv \alpha } ,
\label{EqdRe1b}
\ee
where $\dd H$ has been expressed explicitly.
The stipulation that curvature is negligible implies $H^2\approx 8\pi \rho/3$, hence
\be
\dd R_e \simeq - \frac{1}{2} a \int _a ^{\infty} { \frac{f\dd \rho }{\alpha ^2  H \rho} \drv \alpha } ,
\label{EqdRe1c}
\ee
Recall from Eq.(\ref{Eqconst}) that $f\dd\rho/\rho$ is constant, and may be therefore extracted from the integral.  
The expression for $\dd R_e$ in Eq.(\ref{EqdRe1c}) is thus reduced to its final form
\be
\dd R_e \simeq - \frac{\dd\rho}{2\rho} R_e ,
\label{EqdRe}
\ee
where explicit reference to $t_v$ has been omitted for generality.

We may now determine $T_e$.  It follows from Eqs.(\ref{EqTe0}) and (\ref{EqdRe}) that the temperature of the CEH in a flat FRW universe is given by
\be
T_e \simeq \frac{E_e}{S_e},
\label{EqTe}
\ee
in direct analogy to the temperature of the CAH.
Note that the right side of Eq.(\ref{EqTe}) is asymptotically identical to the static dS temperature, $1/(2\pi R_e)$.

\section{Dynamical First Law for the CEH \label{FLTE} }
Presumably, the FLT for stationary cosmological horizons represents the special case $\drv V = 0$ of a more general, fully dynamical form of the FLT.
Given that Eq.(\ref{EqFLTsch}) is identical to the conventional FLT, but with the sign of $\dd E$ reversed, it is reasonable to expect that the putative dynamical FLT for cosmological horizons has the form
\be
T\drv S = X\drv V - \drv E ,
\label{EqFLTC}
\ee
for some appropriately defined $X$ that is analogous to the quantity of pressure appearing in the conventional FLT.

The purpose of this Section is to demonstrate that Eq.(\ref{EqFLTC}), with $X=P$, produces a successful representation of a dynamical form of the FLT for the CEH in a flat FRW universe, which may be stated as  
\be
T_e \dot{S}_e \simeq P \dot{V}_e - \dot{E}_e .
\label{EqFLTE}
\ee
With the associated terms expressed explicitly, the left and right sides of Eq.(\ref{EqFLTE}) may be written respectively as
\be
T\dot{S}_e \simeq 
2 HV_e \left(1 - \frac{1}{HR_e} \right) 
\left(\rho - \frac{3k}{8 \pi a^2}  \right)  ,
\label{EqFLTEL}
\ee
and 
\be
P \dot{V}_e - \dot{E}_e \simeq 
2HV_e \left(1 - \frac{1}{HR_e} \right) 
3P + A_e (\rho + P) ,
\label{EqFLTER}
\ee
where $A_e$ is the surface area of the CEH.  In the remainder of this Section, the validity of Eq.(\ref{EqFLTE}) is tested in the primary eras associated with the standard FRW cosmology.

During an era of radiation-dominance (RD) we have $\rho \simeq \epsilon$, $P\simeq \rho /3$ and $\dot{R}_e \gg 1$.  If $k/a^2$ is very small in comparison to $\rho$ then the difference between the right and left sides of Eq.(\ref{EqFLTE}) is vanishing during RD.   

Next, consider some late time during vacuum-dominance (VD), when $\mu$ is small with respect to $\rho _{\Lambda}$ and $\epsilon$ is negligible in comparison to $\mu$.  The contribution from $k/a^2$ is assumed also to be negligible in the de Sitter era.  The corresponding cosmic pressure is $P\simeq -\rho_{\Lambda}$, and the Hubble parameter is
\be
H \simeq H_{\Lambda} \left( 1 + \frac{\mu}{\rho_{\Lambda} }  \right)^{1/2} ,
\label{EqHVD}
\ee
where
\be
H_{\Lambda} \equiv \frac{8 \pi}{3}\rho _{\Lambda} =\frac{1}{R_{\Lambda} ^2}.
\ee
With $\mu(\alpha) = \mu (a)a^3 /\alpha ^3$, Eq. (\ref{EqRe}) produces 
\be
R_e \simeq 
R_{\Lambda} - 
\frac{ R_{\Lambda} \mu (a) a^4  }{2\rho _{\Lambda}} \int _a ^{\infty} { \frac{\drv \alpha}{\alpha ^5 } } ,
\label{EqReVD1}
\ee
and thus
\be
R_e \simeq R_{\Lambda} \left( 1 - \frac{\mu}{8 \rho _{\Lambda} }\right).
\label{EqRemu}
\ee
Consequently, we have 
\be
\dot{R}_e =R_e H -1 \simeq \frac{3 \mu}{8 \rho _{\Lambda} }.
\label{EqdotREVD}
\ee
It follows from Eqs.(\ref{EqHVD}) through (\ref{EqdotREVD}) that the right and left sides of Eq.(\ref{EqFLTE}) both reduce to $\pi \mu R_e ^2$ during VD in late times.

\section{Discussion\label{Disc}}
Relying only on first principles, the analysis in Section \ref{Temp} produced an expression for the temperature $T_e$ of the CEH in a flat FRW universe.  
Section \ref{FLTE} introduced a dynamical form of the FLT for the CEH, with the sign of $\drv E$ reversed analogously to the FLT for stationary cosmological horizons \cite{Cai2002a}, \cite{Cai2002b}.
With $T_e$ given by Eq.(\ref{EqTe}), the FLT in Eq.(\ref{FLTE}) is valid during RD and during the late de Sitter era of VD in a flat FRW universe.  

In contrast to the usual assumption that $T_e$ is always given by $(2\pi R_e)^{-1}$, and therefore generally much smaller than unity, the present thesis specifies a temperature that may be larger than unity by many orders.  For instance, consider the standard $\Lambda$-CDM cosmology with an early era of vacuum-driven inflation lasting for roughly 60 $e$-folds.  Although $T_e$ is not generally valid in curved FRW spacetimes, it is valid at the end of inflation when curvature is negligible.
Using the nominal parameters $\Lambda \sim 10^{-122}$, $T_{\phi} \sim 10^{-4}$ and $T_P \sim 10^{-1}$, where $T_P$ is the photon temperature at the Planck time, it follows from Eq. (\ref{EqRe}) that $R_e$ may be as large as $(\Lambda /\rho_{\phi} )^{1/4} \sim 10^{33}$ at the end of inflation \cite{BF2003}.  The corresponding enclosed energy would be order near $10^{86}$, and it follows from Eq.(\ref{EqTe}) that $T_e \sim 10^{20}$ at the end of inflation.  Alternatively, in the absence of inflation, we have $R_e \sim 10^{29}$ and $T_e \sim 10^{29}$ near the Planck era.  The occurrence of such enormous primordial horizon temperatures deserves further consideration.  It is sufficient here to mention that the preponderance of the enclosed fluid is not generally in thermal equilibrium with the horizon temperature. 

If the present model is meaningful then the thermodynamic evolution it specifies for the CEH should be presumably consistent with the Generalized Second Law of Thermodynamics (GSLT).
Such a determination is beyond the scope of this present work.  It is nonetheless appropriate to survey the current understanding of the evolution of the CEH with respect to the GSLT in order to anticipate broadly how Eqs.(\ref{EqTe}) and (\ref{EqFLTE}) may be relevant.

Throughout the evolution of a FRW universe, background particles comprising the cosmic fluid are continuously transported across the CEH, from the interior to the exterior, as a natural consequence of cosmic expansion.  
The continuous migration increases $S_e$, but decreases the entropy $I_e$ of the enclosed fluid.  
In order for the CEH to evolve in a manner that preserves the GSLT, the change in the total entropy $\H _e \equiv S_e + I_e$ during some infinitesimal interval $\dd t$ must be generally non-negative.
If $I_e = V_e \sigma$, where $\sigma$ is the average entropy density of the cosmic fluid, then $\dot{\H}_e \ge 0$ in a broad range of plausible spacetime scenarios, and to that extent the evolution of the CEH conforms to the GSLT \cite{Dav1987}, \cite{DDL2003}.
A more detailed requirement follows from evaluating $\dot{I}_e$ in terms of the Gibbs equation.  
Specifically, as particles comprising the cosmic fluid approach the CEH, they presumably interact with the Hawking radiation emanating therefrom.  If the fluid is locally in thermal equilibrium with the horizon then the Gibbs equation specifies that \cite{IP2006}
\be
\dot{I}_e = \frac{P \dot{V}_e + \dot{E}_e }{T_e} .
\label{EqGE}
\ee
If $\dot{I}_e$ is given by Eq.(\ref{EqGE}), and if $T_e =1/(2\pi R_e)$, then the thermodynamic behavior of the CEH in a vacuum-dominated universe may violate the GSLT \cite{WGA2007}, \cite{WGA2006}.  

According to the present thesis, however, the temperature of the CEH is not generally equal to $T_e =1/(2\pi R_e)$, and the conclusions in Refs.\cite{WGA2007} and \cite{WGA2006} may be substantially different with $T_e$ given by Eq.(\ref{EqTe}).  
Moreover, even if the cosmic fluid is locally in thermodynamic equilibrium with the CEH, the implementation of the Gibbs equation is sensitive to the nature of dark energy.  In fact, in the scenario of holographic dark energy described in Ref. \cite{Li2004}, the asymptotic CEH actually conforms to the GSLT, with $\dot{I}_e$ given by the appropriate Gibbs equation and $T_e = 1/(2\pi R_e)$ \cite{BGLeD2007}.  The degree to which the CEH conforms to the GSLT in the various models of dark energy, with $T_e$ given by Eq.(\ref{EqTe}), is reserved for future study.

\bibliography{thermoCEH}

\end{document}